\documentclass[aps,twocolumn,superscriptaddress]{revtex4}
\usepackage{graphicx}

\begin{document}

\title{Structural and magnetic phase transitions in Na$_{1-\delta}$FeAs}

\author{Shiliang Li}
\email{slli@aphy.iphy.ac.cn}
\affiliation{
Beijing National Laboratory for Condensed Matter Physics and 
Institute of Physics, Chinese Academy of Sciences, P.O. Box 603, Beijing 100190, China
}
\affiliation{
Department of Physics and Astronomy, The University of Tennessee, Knoxville, Tennessee 37996-1200, USA
}
\author{Clarina de la Cruz}
\affiliation{
Department of Physics and Astronomy, The University of Tennessee, Knoxville, Tennessee 37996-1200, USA
}
\affiliation{
Neutron Scattering Science Division, Oak Ridge National Laboratory, Oak Ridge, Tennessee 37831-6393, USA
}
\author{Q. Huang}
\affiliation{
NIST Center for Neutron Research, National Institute of Standards and Technology, Gaithersburg, MD 20899, USA
}

\author{G. F. Chen}
\affiliation{
Department of Physics, Remin University of China, Beijing 100872, China
}
\affiliation{
Beijing National Laboratory for Condensed Matter Physics and 
Institute of Physics, Chinese Academy of Sciences, P.O. Box 603, Beijing 100190, China
}
\author{T.-L. Xia}
\affiliation{
Department of Physics, Remin University of China, Beijing 100872, China
}
\author{J. L. Lou}
\affiliation{
Beijing National Laboratory for Condensed Matter Physics and 
Institute of Physics, Chinese Academy of Sciences, P.O. Box 603, Beijing 100190, China
}
\author{N. L. Wang}
\affiliation{
Beijing National Laboratory for Condensed Matter Physics and 
Institute of Physics, Chinese Academy of Sciences, P.O. Box 603, Beijing 100190, China
}
\author{Pengcheng Dai}
\email{daip@ornl.gov}
\affiliation{
Department of Physics and Astronomy, The University of Tennessee, Knoxville, Tennessee 37996-1200, USA
}
\affiliation{
Neutron Scattering Science Division, Oak Ridge National Laboratory, Oak Ridge, Tennessee 37831-6393, USA
}
\affiliation{
Beijing National Laboratory for Condensed Matter Physics and 
Institute of Physics, Chinese Academy of Sciences, P.O. Box 603, Beijing 100190, China
}

\begin{abstract}
We use neutron scattering to study the spin and lattice structures of 
single crystal and powder samples of Na$_{1-\delta}$FeAs ($T_c = 23$ K). 
On cooling from room temperature, the system goes through a series of phase transitions: first changing the crystal symmetry from tetragonal to orthorhombic at 49 K, then ordering antiferromagnetically with a spin structure similar to that of LaFeAsO and a small moment (0.09$\pm$0.04 $\mu_B$), and finally becoming superconducting below about 23 K. These results confirm that antiferromagnetic order is ubiquitous for the parent compounds of the iron arsenide superconductors, and suggest that the separated structural and magnetic phase transition temperatures are due 
to the reduction in the $c$-axis exchange coupling of the system.   
\end{abstract}


\maketitle
Determining the universal features of iron arsenide superconductors is an important first step in developing a microscopic theory to understand the high-transition temperature (high-$T_c$) superconductivity in these materials \cite{kamihara}.  From the outset, 
it was known that antiferromagnetic (AF) order is a competing ground state to superconductivity in iron arsenide superconductors.  The parent compound LaFeAsO exhibits a structural distortion at 155 K and then orders antiferromagnetically below 137 K,  electron doping to 
induce superconductivity suppresses both the structural distortion and static AF order \cite{cruz,mcguire}.  Although subsequent neutron scattering experiments on the parent compounds of other iron-based superconductors including $R$FeAsO ($R=$ Ce, Pr, Nd) \cite{jzhao1,kimber,jzhao2,ychen}, $A$Fe$_2$As$_2$ ($A =$ Ba, Sr, Ca) \cite{qhuang,jzhao3,goldman}, and FeTe\cite{wei,li1} have found similar lattice distortion and AF order as that of the LaFeAsO, the $M$FeAs system ($M =$ Li, Na) seemed to be an exception to this universal picture
since the initial neutron and X-ray scattering 
experiments have found no evidence of lattice distortion or static AF order 
\cite{cqjin,pitcher,tapp,pwchu}. These results are in contrast with local density approximation (LDA) 
calculations, where the Fermi surfaces and magnetic orders of $M$FeAs
are expected to be similar to that of the LaFeAsO \cite{dsingh,jishi}.  Although recent transport and heat 
capacity measurements on single crystals of Na$_{1-\delta}$FeAs showed two anomalies at 52 K and 41 K that are assigned to structural and AF phase transitions, respectively \cite{gfchen}, muon-spin rotation experiments ($\mu$SR) confirmed only the AF ordering and neutron scattering found 
no evidence for structural distortion \cite{parker2}.

In this paper, we report neutron scattering investigation of spin and lattice structures of single crystals and polycrystalline materials of Na$_{1-\delta}$FeAs. We 
identify a tetragonal to orthorhombic structural phase transition near 50 K
and thus confirm the transport measurements \cite{gfchen}.
Although our neutron powder diffraction measurements were unable to detect AF order due to the small Fe moment, single crystal experiments using thermal triple-axis spectroscopy unambiguously confirmed an AF phase transition
below 40 K and showed that the system forms a collinear in-plane AF spin structure identical to other iron arsenides \cite{jzhao1,kimber,jzhao2,ychen,qhuang,jzhao3,goldman} but doubling the unit cell along the $c$-axis [Fig. 1(a)].  The ordered moment is by far the smallest in iron arsenides, being $0.09 \pm 0.04$ $\mu_{\rm B}$.  
These results suggest that AF order with a collinear in-plane spin structure is ubiquitous property of the
parent compounds of iron arsenide superconductors. We argue that the separated structural and magnetic phase transitions in Na$_{1-\delta}$FeAs is due to reduced $c$-axis exchange coupling and Na deficiencies. 
Bulk superconductivity in Li$_{1-\delta}$FeAs and Na$_{1-\delta}$FeAs can only arise with enough self-doping induced by alkali metal deficiencies.

\begin{figure}
\includegraphics[scale=.4]{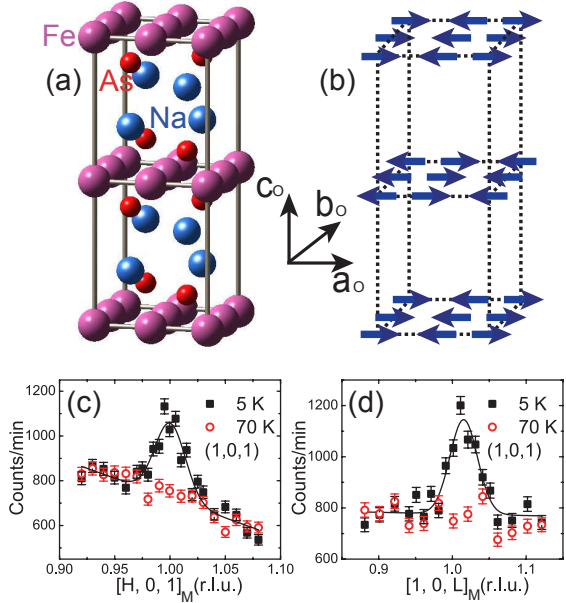}
\caption{
(a) Nuclear and (b) magnetic structures of the ideal NaFeAs. (a) includes two orthorhombic nuclear unit cells for comparison with the magnetic unit cell in (b). 
(c) and (d) show the $[H,0,1]$ and $[1,0,L]$ scans around $(1,0,1)_M$ magnetic Bragg peak at 5 K and 70 K. 
Both peaks disappear at 70 K, indicating their magnetic nature.
}
\label{fig1}
\end{figure}

We prepared 7 grams of polycrystalline Na$_{1-\delta}$FeAs sample and 0.6 gram of single crystals as described in Ref. \cite{gfchen}. The resistivity measurement gives the onset and zero-resistivity $T_c$ as 23 K and 8 K, respectively \cite{gfchen}. Powder neutron diffraction measurements were performed on the BT-1 high resolution powder diffractometer at the NIST Center for Neutron Research. The BT-1 diffractometer has a Ge(3,1,1) monochromator ($\lambda = 2.0785$ {\AA}) and collimators with horizontal divergences of $15^{\prime}$-$20^{\prime}$-$7^{\prime}$ full width at half maximum (FWHM). Powder diffraction data refinement was done by using the GSAS program. The measurements on single crystals were carried out on the HB-1 triple-axis spectrometer at the High Flux Isotope Reactor, Oak Ridge National Laboratory. We fixed final neutron energy at $E_f = 13.5$ meV and used PG(0,0,2) (pyrolytic graphite) as monochromator and analyzer. The collimations for magnetic and structural measurements are $48^\prime$-$60^\prime$-$80^\prime$-$120^\prime$ and $15^\prime$-$20^\prime$-$20^\prime$-$30^\prime$, respectively. We denote $Q$({\AA}$^{-1}) = (2\pi H/a, 2\pi K/b, 2\pi L/c)$, where $a_T = b_T = 3.94481(3)$ {\AA}, $c_T$ = 6.99680(8) \AA\ for tetragonal structure and $a_{O} = 5.58906(8)$ \AA, $b_{O} = 5.56946(8)$ \AA, $c_{O} = 6.9919(1)$ \AA\ for orthorhombic structure. The magnetic unit cell is defined as $a_{O}\times b_{O}\times 2c_{O}$.

\begin{table}
\caption{\label{refinement}Refinement results of powder diffraction data}
\begin{ruledtabular}
\begin{tabular}{lccccc}
\multicolumn{6}{l}{Na$_{0.985}$FeAs(5 K), $Cmma$, $\chi^2$ = 1.453} \\
\multicolumn{6}{l}{$a_O = 5.58906(8)$ {\AA}, $b_O = 5.56946(8)$ {\AA}, $c_O = 6.9919(1)$ {\AA}}\\
\hline
Atom & site & x & y & z & occupancy \\
Na & 4a & 0 & 0.25 & 0.3533(2) & 0.985(7)\footnote{Na occupancy is calculated as the mean value of those at several temperatures.} \\
Fe & 4g & 0.25 & 0 & 0 & 1\footnote{Fe and As occupancies are fixed to 1.} \\
As & 4a & 0 & 0.25 & 0.7977(1) & 1\footnotemark[2] \\
\multicolumn{6}{l}{Selected bond lengths and angles:} \\
\multicolumn{3}{l}{Fe-Fe$\times$2 2.79453(4) \AA} & \multicolumn{3}{l}{Fe-Fe$\times$2 2.78473(4) \AA} \\
\multicolumn{3}{l}{Fe-As$\times$4 2.4272(4) \AA} & \multicolumn{3}{l}{Na-As$\times$1 3.107(2) \AA} \\
\multicolumn{3}{l}{Na-As$\times$2 2.9874(6) \AA} & \multicolumn{3}{l}{Na-As$\times$2 2.9782(6)}\\
\multicolumn{3}{l}{Fe-As-Fe$\times$2 70.29(1)$^{\circ}$} & \multicolumn{3}{l}{Fe-As-Fe$\times$2 70.01(1)$^{\circ}$} \\ \multicolumn{3}{l}{Fe-As-Fe$\times$2 108.72(3)$^{\circ}$} & & & \\
\hline 
\multicolumn{6}{l}{Na$_{0.985}$FeAs(70 K), $P4/nmm$, $\chi^2$ = 1.685}\\
\multicolumn{6}{l}{$a_T = b_T = 3.94481(3)$ {\AA}, $c_T = 6.99680(8)$ {\AA}}\\
\hline
Atom & site & x & y & z & occupancy \\
Na & 2c & 0.25 & 0.25 & 0.3535(2)  \\
Fe & 2a & 0.75 & 0.25 & 0 & 1\footnotemark[2]   \\
As & 2c & 0.25 & 0.25 & 0.7976(1) & 1\footnotemark[2]  \\
\multicolumn{6}{l}{Selected bond lengths and angles:} \\
\multicolumn{3}{l}{Fe-Fe$\times$4 2.78941(3) \AA} & \multicolumn{3}{l}{Fe-As$\times$4 2.4282(4) \AA} \\
\multicolumn{3}{l}{Na-As$\times$4 2.9830(6) \AA} & \multicolumn{3}{l}{Na-As$\times$1 3.107(2) \AA} \\
\multicolumn{3}{l}{Fe-As-Fe$\times$4 70.11(2)$^{\circ}$} & \multicolumn{3}{l}{Fe-As-Fe$\times$2 108.64(3)$^{\circ}$}\\
\end{tabular}
\end{ruledtabular}
\end{table}

\begin{figure}
\includegraphics[scale=.5]{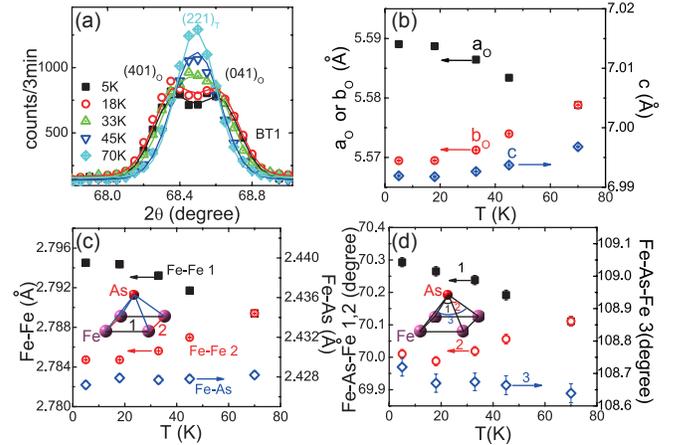}
\caption{(a)
Temperature dependence of the 2$\theta$ scans around the $(2,2,1)_T$ Bragg peak 
in powder diffraction data across the tetragonal to orthorhombic phase transition.
Clear splitting of the peaks is seen below 33 K.  We collected full powder diffraction
spectra at several temperatures 
and refinement results gave the temperature dependence of several key parameters: 
(b) lattice constants $a$, $b$ and $c$; (c) bond lengths of Fe-Fe and Fe-As; (d) bond angles of Fe-As-Fe. In all panels, left and right axes have the same scale. 
}
\label{fig2}
\end{figure}

We first describe neutron powder diffraction measurements. 
At a temperature well above the structural and magnetic phase transitions ($T=70$ K), 
Rietvelt analysis reveals a tetragonal structure with space group $P4/nmm$ 
consistent with earlier results \cite{parker2}. After cooling the sample down to 5 K, the tetragonal $(2,2,1)$ peak splits into two peaks as shown in Fig. 2(a).  Detailed temperature dependence of the 
$(2,2,1)$ profiles in Fig. 2(a) reveals that the 
tetragonal to orthorhombic structural phase transition occurs near 45 K.  Refinement of the diffraction pattern supports an orthorhombic structure at low temperature and detailed structural parameters are listed in Table I for the two temperatures investigated.  Fixing the occupancies of Fe and As to 1, we obtain the Na content as 0.985(7), or equivalently 1.5\% Na deficiencies. 
Figures 2(b)-2(c) show the temperature dependence of some key parameters.  As a function of increasing temperature,
the Fe-Fe bond length decreases until they become equal (tetragonal) while the Fe-As distance remains unchanged [Fig.1(c)]. The orthorhombic lattice parameters $a_O$ and $b_O$ 
behave similarly [Fig. 1(b)].  While
the nearest neighbor bond angles change in opposite directions with increasing temperature, the diagonal bond angle
is essentially temperature independent [Fig. 2(d)].  These results are similar to those of LaFeAsO \cite{nomura}, and thus suggesting the same underlying mechanism for the structural phase transition. 

\begin{figure}
\includegraphics[scale=.7]{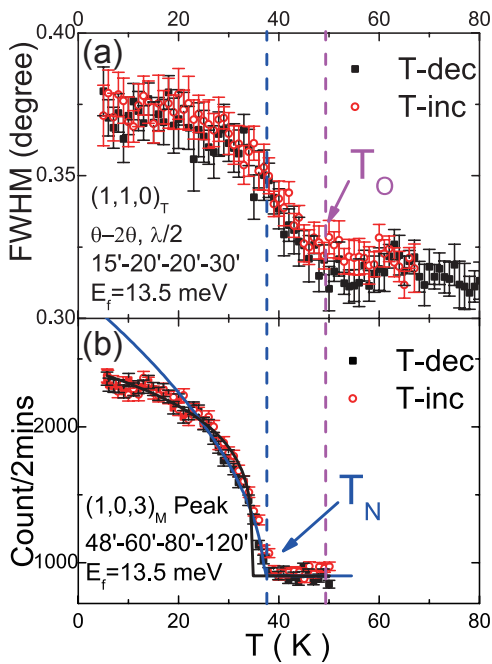}
\caption{
(a) Temperature dependence of the FWHM of $\theta$-$2\theta$-scan at the nuclear peak (1,1,0)$_T$ position using $\lambda/2$ scattering by removing the pyrolytic graphite (PG)  filter. The peak width clearly increases below about 50 K as marked by the vertical line.
(b) Temperature dependence of the peak intensity at the AF peak $(1,0,3)_M$ position suggesting
a N$\rm \acute{e}$el temperature of 39 K. The black line is the fitted result for the whole temperature range by using a simple Ising model described in the text, while the blue line only focuses on temperatures above 20 K.
No anomaly is seen across $T_c$, suggesting that superconductivity is filamentary and not a bulk phenomenon.
}
\label{fig3}
\end{figure}

To precisely determine the structural transition temperature, we carefully measured the temperature dependence of the $(1,1,0)_T$ peak width (in FWHM) on the single crystal sample using $\lambda/2$
as shown in Fig. 3(a). Although the resolution of the triple-axis spectrometer is not good enough to resolve two
separate peaks from the $\theta-2\theta$ scans at low temperature, the change of the FWHM reveals a 
structural phase transition temperature of $T_{O} \approx$ 50 K, which is consistent with the higher transition temperature in the transport measurements \cite{gfchen}.

To search for possible magnetic order, we carried out measurements using triple-axis spectroscopy on both the powder samples and single crystals.  While we cannot find any magnetic peak in the powder diffraction data due to small Fe moment, we observe clear AF order on single crystals at low temperatures. 
It turns out that the in-plane AF unit cell of Na$_{1-\delta}$FeAs is identical to that of LaFeAsO \cite{cruz}, where the 
magnetic Bragg peaks are observed at $(1,0,L)_M$ ($L = 1,3,5,7$).  Figures 1(c) and 1(d) show 
wave vector scans along the orthorhombic $[H,0,1]$ and $[1,0,L]$ directions \cite{jzhao2} at 5 K and 70 K.
The resolution limited peaks around $(1,0,1)_M$ at 5 K disappear at 70 K [Figs. 1(c) and 1(d)].
Figure 3(b) shows the temperature dependence of the scattering at wave vector $Q=(1,0,3)_M$, where we estimate that 
 the onset magnetic transition temperature is about 37 K with less than 1 K thermal hysteresis [Fig. 3(b)].  These results are consistent with transport measurements where the $\sim$40 K transition is identified as magnetic in nature.
In a simple Ising model, the magnetic order parameter is related to temperature via 
$\phi(T)^2 \propto (1-T/T_N)^{2\beta}$. For
BaFe$_2$As$_2$, AF phase transition has a critical exponent $\beta=0.103\pm0.018$ \cite{wilson}.
Fitting the whole temperature dependence of (1,0,3)$_M$ peak intensity yields an unreasonable $T_N = 34.7 \pm 0.9$ K, shown as the black line in Fig. 3(b). Limiting the fitting range to temperatures above 20 K 
gives a $T_N = 37.1\pm0.2$ K and $\beta = 0.28\pm 0.02$, considerably higher than that of BaFe$_2$As$_2$.  However, we caution that the order parameter was measured using peak intensity on single crystals, rather than integrated intensity measurement.

A collinear AF structure could either have spin directions along the $a$- or $b$-axis, which would correspond to 
 magnetic peaks at $(1,0,L)_M$ or $(0,1,L)_M$ directions, respectively. In previous work on SrFe$_2$As$_2$ \cite{jzhao3}, the AF ordering direction was determined  to be along the $a$-axis direction by comparing the $(1,0,1)$ magnetic bragg peak with $\lambda/2$ scattering from $(2,0,1)$ and $(0,2,1)$ nuclear Bragg peaks.
Since the orthorhombic peak splitting in the case of Na$_{1-\delta}$FeAs is rather small, we used very tight collimations for this purpose.  Figure 4(a) shows the magnetic Bragg peak near $(1,0,L)_M$ or $(0,1,L)_M$ and
$\lambda/2$ scattering from twinned peaks of $(1,0,0.5)$ and $(0,1,0.5)$.  It is immediately clear that 
the magnetic (1,0,1)$_M$ peak position in $\theta-2\theta$ scan is smaller than that of the nuclear peak.
This result is consistent with earlier work on SrFe$_2$As$_2$ \cite{jzhao3}.

\begin{figure}
\includegraphics[scale=.7]{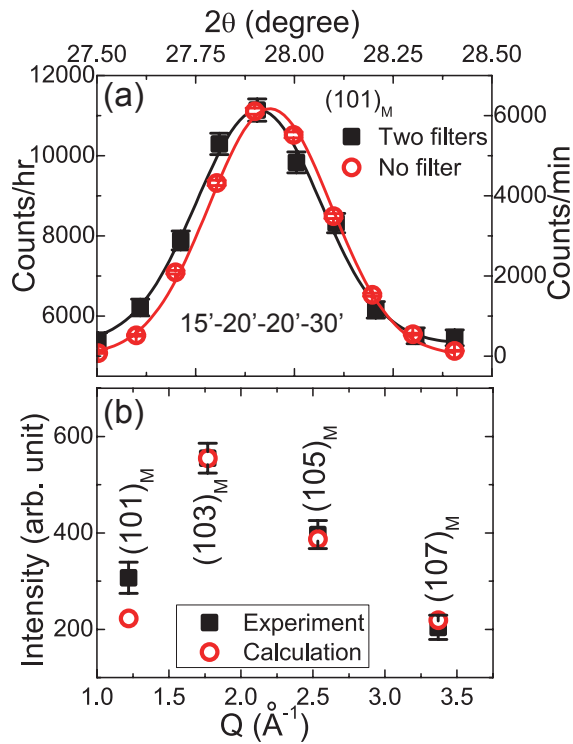}
\caption{
(a) Comparison of the $\theta$-$2\theta$ scans at $(1,0,1)_M$ with and without PG filter, measuring magnetic and nuclear peaks, respectively. 
(b) Calculated and experimental intensities of magnetic peaks $(1,0,L)_M$ ($L = 1, 3, 5, 7$). In the calculation, the moment is assumed to be along $a_{O}$ axis.
}
\label{fig4}
\end{figure}

To determine the spin direction, we calculate the magnetic structure factors by assuming that the moments 
point to the $a$-axis direction. 
The observed magnetic intensities are obtained by integrating the $\theta$-$2\theta$ scans at the expected magnetic peak positions $(1,0,L)$ ($L=1,3,5,7$) in the three-axis mode. A comparison of the calculated and observed magnetic peak intensities shown in Fig. 4(b) reveals that such a model explains the data reasonably well.
The small deviation between the observed and calculated intensities sets a limit of the moment direction to be within 15 degrees away from the $a$-axis. Therefore, the magnetic structure in Na$_{1-\delta}$FeAs is the same as that 
in the $A$Fe$_2$As$_2$ system as shown in Fig. 1(b) \cite{qhuang,jzhao3,goldman}. Assuming this spin structure, 
we can estimate an Fe moment of $0.09 \pm 0.04$ $\mu_B$ by comparing intensities of the AF Bragg peaks with with a series of nuclear peaks. Such a small moment explains why we as well as another group \cite{parker2} 
cannot find any AF Bragg peaks in neutron powder diffraction measurements.

Our results on Na$_{1-\delta}$FeAs suggest that orthorhombic structure 
and collinear AF spin ordering in Fig. 1(b) are ubiquitous properties of 
undoped iron arsenides. Because  of the difficulty in making stoichiometric samples of NaFeAs,
Na deficiencies in the as-grown NaFeAs dopes holes onto the FeAs-plane that can induce superconductivity \cite{gfchen}.  For BaFe$_2$As$_2$, transport and neutron scattering experiments have shown 
that electron-doping reduces the $c$-axis magnetic exchange coupling and separates the structural and magnetic phase transitions \cite{ni,chu,lester,harriger}. Theoretically, it has been argued that the strength of the $c$-axis magnetic coupling controls the simultaneous or separated structural/magnetic phase transitions \cite{hu}.
For lightly doped BaFe$_{1.96}$Ni$_{0.04}$As$_2$, inelastic neutron scattering experiments showed a dramatic drop in the $c$-axis correlations with electron-doping.  Based on density function theory calculations, the $c$-axis exchange coupling of NaFeAs is smaller than that of BaFe$_2$As$_2$ but larger than that of LaFeAsO.
Experimentally, the temperature separations between structural and magnetic phase transitions are 
similar for LaFeAsO \cite{cruz} and NaFeAs. In addition, the AF order parameter showed no anomaly across $T_c$ similar to those of BaFe$_{1.96}$Ni$_{0.04}$As$_2$ [Fig. 3(b)].
These results suggest that superconductivity in our Na$_{1-\delta}$FeAs is filamentary and not a bulk phenomenon.
Since our single crystals of Na$_{1-\delta}$FeAs is slightly doped away from
ideal stoichiometry (Na deficiency), it is unclear whether the observed small moment and large
differences in structural/magnetic phase transition temperatures are the intrinsic or doping-induced 
$c$-axis coupling reduction.  Future inelastic neutron scattering experiments will be able to determine the 
exchange coupling along the $c$-axis.

In conclusion, we have determined the lattice and magnetic structures of single crystal Na$_{1-\delta}$FeAs.  Our results indicate that the parent materials of NaFeAs and LiFeAs superconductors have orthorhombic lattice 
distortion and collinear AF order.  This work establishes 
that the orthorhombic structure and AF collinear order are ubiquitous to all 
undoped iron arsenide materials.  Superconductivity arises from electron or hole-doping of their AF parent compounds, and therefore suggesting that spin fluctuations are important for superconductivity of these materials.

We thank Jiangping Hu, J. A. Fernandez-Baca, Tao Xiang, Zhong-Yi Lu for helpful discussions. 
This work is supported by the U.S. NSF No. DMR-0756568, U.S. DOE BES No.
DE-FG02-05ER46202, and by the U.S. DOE, Division of Scientific
User Facilities. The work in IOP is supported by
the Ministry of Science and Technology of China and NSFC.

\end{document}